\documentstyle[preprint,prl,aps]{revtex}

\begin{document}
\draft
\preprint{RU9542}
\title{Heated nuclear matter, condensation phenomena and the hadronic
equation of state}
\author{K. C. Chase and A. Z. Mekjian}
\address{Department of Physics, Rutgers University \\
Piscataway, New Jersey 08854}
\date{\today}

\maketitle

\begin{abstract}
The thermodynamic properties of heated nuclear matter are explored
using an exactly solvable canonical ensemble model.
This model reduces to the results of an ideal Fermi gas at low temperatures.
At higher temperatures, the
fragmentation of the nuclear matter into clusters of nucleons
leads to features that resemble a Bose gas.
Some parallels of this model with the phenomena of Bose condensation
and with percolation phenomena are discussed.
A simple expression for the hadronic equation of state is obtained from
the model.
\end{abstract}

\narrowtext


Properties of heated nuclear matter are of current interest, with
heavy-ion collisions used to obtain information about nuclear
matter away from the typical nuclear matter density and temperature.
Theoretical concerns have centered around the nuclear equation of
state, the specific heat, the behavior of the entropy per particle,
and other thermodynamic issues~\cite{Bertsch,Csernai,DasGupta}.
The notion that nuclear matter undergoes a phase transition has gained
special attention~\cite{Gilkes}.
Most treatments of the equation of
state have treated heated nuclear matter as a homogeneous system.
Statistical~\cite{Jaqaman1,Jaqaman2,DeAngelis1,DeAngelis2,Gross1,Gross2,Sobotka,Aichelin,Koonin,Boal,Cole,Fai,Friedman,Bondorf,Sneppen,Mekjian1,Mekjian2,Mekjian3,Chase1}
and percolation~\cite{Campi,Desbois,Bauer}
models have been used by many groups to study nuclear
multifragmentation and the role of clusters on nuclear properties.
This paper explores various thermodynamic properties using an exactly
solvable canonical ensemble model that allows for clusterization
and leads to simple analytic expressions for thermodynamic quantities
such as the equation of state.
This canonical ensemble approach has properties that are similar in
many ways to those obtained from models used in other areas, such as
Bose condensation, Feynman's approach to the $\lambda$ transition in
liquid helium~\cite{Feynman} and in polymer physics~\cite{Kelly}.
For example, the cycle class decomposition of the symmetric group which
appears in the symmetry of the bosonic system in Feynman's approach
is isomorphic to the
cluster structure of a heated nuclear system.  The cost function of
moving a helium atom from one location to another is the same as
internally exciting a cluster.

The canonical ensemble partition function given below is obtained from
the following simple picture.  At $T=0$, a system of $A$ nucleons is
in its ground state, which is treated as a degenerate Fermi gas.  As
$T$ increases, the Fermi gas is excited into low lying excited states
whose level density grows as $\rho(E) \sim \exp (2 \sqrt{a E})$.
Particles and clusters can also be emitted with increasing $T$.
The formation of clusters is an expression of the attractive nature of
the nucleon-nucleon interaction.  The repulsive nature of this
interaction will be treated here in terms of a density dependent
Skyrme approach~\cite{Cortin,Jaqaman3}.  At high density an excluded
volume correction is needed to avoid some unphysical consequences of
the model.  At very large $T$ clusterization is unlikely, and the
nucleons behave essentially as a dilute Maxwell-Boltzmann gas.

Considered as a statistical system, this model exhibits a phase
transition in the fixed density but infinite $A$ limit, at which the
specific heat per nucleon is maximal.
An infinite cluster suddenly appears at this point
in a manner similar to the infinite cluster of percolation theory.
Real nuclear systems are far from infinite collections of nucleons,
but the evidence of such phase transitions can be obtained from a
finite scaling analysis in the region of the phase
transition~\cite{Campi,Gilkes}.


In earlier papers~\cite{Mekjian1,Mekjian2,Mekjian3,Chase1}
the question of the thermodynamics
of a fragmenting system was raised.  There, it was assumed that the
thermodynamic variables were contained in a single
parameter $x$ such that the weight of a particular cluster partition
was proportional to $x^{m}$, where $m$ is the total number of
clusters in that partition.  Specifically, the partition function of
the system was a polynomial in the variable $x$ of degree $A$, where
$A$ is the number of particles in the system.  This $x$ contains
physical quantities as discussed below in Eq.~(\ref{eq:model-xy}).
The computation of various thermodynamic
quantities could then be reduced to functions of $x$ and
various moments of the number of clusters.
Here the slightly more general case where there are two
thermodynamic parameters $x$, $y$ with the probability of a particular
fragmentation outcome proportional to $x^{m} y^{A-m}$ is analyzed.
This choice reduces to the early models by noting
$x^{m} y^{A-m} = y^{A} (x/y)^{m}$ where $x/y$ can be identified with
the tuning parameter $x$ used earlier and $y^{A}$ is an overall
factor which has no effect on cluster yields.  However, this more
general case is necessary for an evaluation of some thermodynamic
properties.

We begin by assuming that each fragmentation outcome happens with a
probability proportional to
\begin{equation}
W(\vec{n}) = \prod_{k \ge 1} {x_{k}^{n_{k}} \over n_{k}!}
   = \prod_{k \ge 1} {1 \over n_{k}!}
     \left( {x y^{k-1} \over \beta_{k}} \right)^{n_{k}} \;,
\label{eq:weight}
\end{equation}
where $x$ and $y$ are functions of the thermodynamic variables $V$,
$T$, $\beta_{k}$ is the cluster size dependence of the weight
discussed below, and $\vec{n} = (n_{1}, n_{2}, \ldots)$ is the
fragmentation vector, with $n_{k}$ the number of fragments with $k$
nucleons such that $\sum_{k} k n_{k} = A$.  The free energy for such a
canonical system is given by $F_{A}(V, T) = - k_{B}T \ln Z_{A}$ with
\begin{eqnarray}
Z_{A}
  & = & \sum_{\vec{n} \in \Pi_{A}} W(\vec{n})
    =   \sum_{m=0}^{A} Z_{A}^{(m)}(\vec{\beta}) x^{m} y^{A-m} \;,
\end{eqnarray}
where $\Pi_{A}$ is the set of all partitions of $A$ and $Z_{A}^{(m)}$
is a function only of the vector $\vec{\beta}$, not of the
thermodynamic variables.

It is fairly straightforward to derive relations
between logarithmic derivatives of the partition functions and moments
of the multiplicity $m = \sum_{k} n_{k}$, e.g.
\begin{equation}
\begin{array}{cc}
x {\partial \over \partial x} \ln Z_{A} = \langle m \rangle, &
\left( x {\partial \over \partial x} \right)^{2} \ln Z_{A}
  = \langle m \rangle_{2},
\end{array}
\label{eq:compute-moments}
\end{equation}
where
$\langle m \rangle_{k} = \langle (m-\langle m \rangle)^{k} \rangle$
is the $k$th central moment.
In general, $(x {\partial \over \partial x})^{k} \ln Z_{A}(x)$ is the
$k$'th cumulant moment of $m$.  The relations for derivatives with
respect to $y$ are similar.

Given these formulae, the determination of thermodynamic quantities
can be reduced to a computation of the cumulant moments of the
fragmentation multiplicity $m$, since only
logarithmic derivatives of the partition function (such as those
given above) are used in the determination of the energy, pressure,
specific heat, etc.  Applying the usual relations between the
thermodynamic functions and the partition function, the result for
the specific heat and pressure is
\begin{eqnarray}
{C_{V} \over k_{B}} & = &
        \langle m \rangle
          \left( {2 T \over x} {\partial x \over \partial T}
            -\left( {T \over x} {\partial x \over \partial T} \right)^{2}
            + {T^{2} \over x} {\partial^{2} x \over \partial T^{2}}
          \right) \nonumber \\
  &   & +(A - \langle m \rangle)
          \left( {2 T \over y} {\partial y \over \partial T}
            -\left( {T \over y} {\partial y \over \partial T} \right)^{2}
            + {T^{2} \over y} {\partial^{2} y \over \partial T^{2}}
          \right) \nonumber \\
  &   & +(\langle m^{2} \rangle - \langle m \rangle^{2})
          \left( {T \over x} {\partial x \over \partial T} -
                 {T \over y} {\partial y \over \partial T}
          \right)^{2} \nonumber \\
{P V \over k_{B} T} & = &
         \langle m \rangle
           {V \over x} {\partial x \over \partial V}
         +(A- \langle m \rangle)
           {V \over y} {\partial y \over \partial V} \;,
\label{eq:therm-funcs}
\end{eqnarray}
where the last equation is the equation of state.
The calculation of the cumulant moments of $m$
can be done by applying Eq.~(\ref{eq:compute-moments}) and using
a recursive evaluation of the coefficients of
the partition function $Z_{A}^{(m)}$,
namely~\cite{Chase1},
\begin{equation}
Z_{A}^{(m)} = {1 \over m}
  \sum_{k=1}^{A} {1 \over \beta_{k}} Z_{A-k}^{(m-1)} \;,
\end{equation}
where $Z_{A}^{(1)} = 1/\beta_{A}$.  The whole recursive procedure is
easy to implement as a computer program.

The choice of $x$, $y$ and $\beta_{k}$ is determined by the physics of
the situation.
The term $x^{m}$ comes from phase space factors and translational
partition function considerations, namely $x = V/\lambda^{d}$,
with $\lambda$ the thermal wavelength of a nucleon
and $d$ the dimensionality of the system (with $d=3$ appropriate
for nuclear fragmentation).
This $x$, with $\beta_{k} = k^{1+d/2}$
in Eq.~(\ref{eq:weight}) is essentially the weight given
to Bose condensation problems in $d$ dimensions.  In this case,
$n_{k}$ is the number of cycles of length $k$ in a cycle class
decomposition of the permutation associated with a particular Bose gas
state.  For a nuclear system, the term $y$ is due to cluster binding
and internal excitations.
In a simplified view of binding energy considerations, each fragment
of size $k$ has a binding energy of $a_{V} (k-1)$, so
a total binding energy of $E_{B} \sim a_{V} (A-\langle m \rangle)$
suggests a Boltzmann weight of $y = \exp(a_{V}/k_{B}T)$.
It is appropriate to also include in this factor a Fermi gas level
density term arising from internal excitations, leading to the result,
\begin{eqnarray}
x & = & V \left( {2 \pi m k_{B} T \over h^{2}} \right)^{d/2} \nonumber \\
y & = & \exp \left\{ {a_{V} \over k_{B} T} +
          {k_{B} T \over \varepsilon_{0}} {T_{0} \over T+T_{0}}
        \right\} \;,
\label{eq:model-xy}
\end{eqnarray}
where $a_{V}$ is the binding energy per nucleon,
$\varepsilon_{0}$ is the level spacing parameter for excited states,
and $T_{0}$ is a temperature cutoff factor for internal excitations.
The expression $x/y$ is the tuning parameter in previous
papers~\cite{Mekjian1,Mekjian2,Mekjian3,Chase1}.

In general the parameters $a_{V}$ and $\varepsilon_{0}$ are density
dependent.  In the Skyrme approach, $a_{V}(\rho)$ is given by
\begin{eqnarray}
a_{V}(\rho)
  & = & a_{D} \left( {\rho \over \rho_{0}} \right)^{2/3} -
        a_{0} \left( {\rho \over \rho_{0}} \right) +
	a_{3} \left( {\rho \over \rho_{0}} \right)^{1+\sigma}
\end{eqnarray}
and for a Fermi gas,
$\varepsilon_{0}(\rho) = (4/\pi^2) \varepsilon_{F}(\rho)$ with
$\varepsilon_{F}(\rho) \sim \rho^{2/3}$ the Fermi energy,
$a_{D} = 3 \varepsilon_{F}(\rho_{0})/5$ and $a_{0}, a_{3}$ are Skyrme
parameters, which can be
determined by requiring the binding energy/nucleon to be a maximum at
$\rho = \rho_{0}$ with the value of $8.0$ MeV.
It should be noted that the factor $\exp (k_{B} T/ \varepsilon_{0})$
can be rewritten as $\exp (a (m d^{2}/2 \hbar^{2}) k_{B} T)$.
Using the above relation between Fermi energy and the density of
states, $a$ is a numerical constant close to $1$ and $d^{3} = V/A$.
This factor with $a = 1$ is the cost function of moving a Helium
atom from one location to another in Feynman's approach to the
$\lambda$ transition in liquid Helium~\cite{Feynman}.

The only remaining parameter to set in this model is $\beta_{k}$.
If the weight $W(\vec{n})$ from Eq.~\ref{eq:weight} was for a Bose
gas, then~\cite{Huang}
\begin{equation}
x_{k} = {x \over k^{\tau}} + {1 \over k} \;,
\end{equation}
with $\tau = 1+d/2$.  The $1/k$ term arises from the zero
momentum states of the Bose condensate.
This additional term is irrelevant at high temperatures,
but of some importance below the critical point.
Since the low temperature behavior we want is essentially Fermi gas
like, we can ignore the zero momentum term, which suggests we use
$\beta_{k} = k^{\tau}$.
Here $\tau$ is the critical parameter $\tau$ discussed
in Fisher~\cite{Fisher}, Finn, et.~al.~\cite{Finn} and Gilkes,
et.~al.~\cite{Gilkes}.  We choose $\tau = 5/2$ to match the exponent of a
Bose gas in three dimensions, but data from~\cite{Gilkes} suggests a
somewhat lower $\tau \approx 2.2$.  Although in a Bose gas $\tau$ is
fixed by the dimension of the system, this nuclear fragmentation model
is not so constrained, and a different $\tau$ can be used.

Applying the above choice of parameters to Eq.~(\ref{eq:therm-funcs})
gives
\begin{eqnarray}
{C_{V} \over k_{B}} = &&
  \langle m \rangle {d \over 2}
  + (A-\langle m \rangle) {2 k_{B} T \over \varepsilon_{0}}
    \left( {T_{0} \over T+T_{0}} \right)^{3}
    \nonumber \\
  &&+ \langle m \rangle_{2}
    \left({d \over 2} + {a_{V} \over k_{B} T}
      - {k_{B} T \over \varepsilon_{0}}
        \left( {T_{0} \over T+T_{0}} \right)^{2}
    \right)^{2} \nonumber \\
P V = && \langle m \rangle \nonumber k_{B} T + (A- \langle m \rangle)
     \nonumber \\
  && \times \left(
    {2 \over 5} \varepsilon_{F}(\rho)
      \left( 1 + {5 \pi^{2} \over 12}
         \left({k_{B} T \over \varepsilon_{F}(\rho)}\right)^{2}
            {T_{0} \over T+T_{0}} \right) \right.
     \nonumber \\
  && \hspace{0.5cm}-\left. a_{0} \left( {\rho \over \rho_{0}} \right)
   +  a_{3} (1+\sigma)
     \left( {\rho \over \rho_{0}} \right)^{1+\sigma} \right) \;.
\end{eqnarray}
Note that the equation of state is particularly simple.
The first term is the ideal gas law, while the second term contains
the low temperature degeneracy pressure and interaction terms from the
Skyrme potential.

The specific heat and equation of state given by the above expression
are plotted in Fig.~\ref{fig:thermodynamic-functions}.
For low $T$, $C_{V} \propto T$ and is that of a heated
nucleus of nucleons treated as a nearly degenerate Fermi gas.  At
higher $T$, nucleons and clusters are emitted from this nucleus,
increasing the number of degrees of freedom and the specific heat.
At a relatively low temperature this effect causes a transition from a
degenerate Fermi gas to a Bose gas-like state.
This change in $C_{V}$ shows up in the figure as a shoulder.
At very high $T$ the nucleus of $A$ nucleons dissolves into $A$
independent nucleons and $C_{V} = {\scriptstyle {1 \over 2}} d k_{B} A$.
The ``cusp'' like behavior of $C_{V}$
can be understood as a critical point in the infinite $A$ limit as
discussed in the next paragraph.
A rounded peak is seen instead of the cusp because of the finite size
of the system.
Figure~\ref{fig:thermodynamic-functions}(b) illustrates the behavior
of $P$ with $V/A$ in the transition region.  In this region, the
equation of state behaves in many ways like a van de Waals gas.


The thermodynamic limit  $A,V \rightarrow \infty$ with
$\rho = A/V$ finite is of interest since it specifies the critical
point behavior and the nature of any phase transitions.
In this case, we can work in the
grand canonical limit and consider the question of behavior of the
largest cluster.  A specific concern
is whether this theory has the same characteristic as the
infinite cluster in percolation theory~\cite{Stauffer}, i.e. for
$p > p_{c}$ the infinite cluster exists while for $p < p_{c}$ it does
not exist.
In the grand canonical limit $\langle n_{k} \rangle$ is given by
\begin{equation}
\langle n_{k} \rangle = {x \over y} {e^{\mu k/k_{B} T} \over k^{\tau}}
  = {x \over y} {z^{k} \over k^{\tau}}
\end{equation}
which implies the mass constraint
\begin{equation}
{A \over x/y}= \sum_{k} z^{k} k^{1-\tau}
\end{equation}
Here $A/(x/y)$ is finite in the thermodynamic limit since
$x \propto V$ and $y = y(\rho, T)$.
For $z \le 1$, the sum is always less than or equal to the
case $z = 1$, for which the sum gives $\zeta(\tau-1)$, i.e. the
Riemann zeta function.
When $z > 1$ the sum diverges and therefore the constraint cannot
hold.  Now $A/(x/y)$ is a function of $A/V, T$ which are fixed, so $z$
is the only parameter we can adjust.  If $(x/y)/A \le 1/\zeta(\tau-1)$ then
we can find a $z \le 1$ such that the mass constraint is satisfied.
However, if $(x/y)/A > 1/\zeta(\tau-1)$ then the
constraint can't be met.  The sum must be truncated, which implies
that the expectation of certain large clusters must be identically zero.
In other words, when $(x/y)/A > 1/\zeta(\tau-1)$, there can be no infinite
cluster, and $(x/y)_{c}/A = 1/\zeta(\tau-1)$ defines
a critical point for this system, which is identical
to the critical point of a Bose gas if $y = 1$.
Since $\zeta(\tau-1) < \infty$ only if $\tau > 2$, this also implies
that the infinite cluster can not exist if $\tau \le 2$.

Universality implies the model near the critical point is specified
uniquely by two critical exponents.  We already mentioned the critical
exponent $\tau$, but another exponent is needed to determine the
critical behavior.  By the above discussion,
the fraction of mass not in the infinite cluster,
$m_{x} = \lim_{A \rightarrow \infty} \sum_{k=1}^{A-1} k \langle n_{k}
\rangle/A$
satisfies $m_{x} < 1$ for $T \le T_{C}$ and
$m_{x} = 1$ for $T>T_{C}$.  Thus
$\langle n_{\infty} \rangle = 1 - m_{x}$ is zero for $T>T_{C}$.
Near $T=T_{C}$, $\langle n_{\infty} \rangle \sim (T-T_{C})/T_{C}$.
The analog of $m_{x}$ in the Bose system is the fraction of Bose
particles in excited states, while $\langle n_{\infty} \rangle$ is the
fraction in the condensed ground state.
For the Bose system the order parameter is taken as the square root of
the number of Bosons in the ground state, and in analogy
$\sqrt{\langle n_{\infty} \rangle} \propto (T-T_{C})^{1/2}$, which gives the
second critical exponent $\beta = 1/2$.


In summary, this paper introduced a simplified model for
nuclear systems which attempts to be valid across a wide range of
temperatures and densities.  Building on a model
developed earlier~\cite{Mekjian1,Mekjian2,Mekjian3,Chase1} with
simple expressions for cluster yields, fluctuations and
correlations useful in the analysis of inclusive and exclusive data,
the thermodynamic properties of the model have been made explicit and
the corresponding thermodynamic functions such as the pressure and
specific heat were shown to be simple analytic functions of the
density, temperature and cumulant moments of the multiplicity.
At typical nuclear matter densities and
very low temperatures the model is equivalent to a nearly degenerate
Fermi gas.  At very high temperatures and low densities it reduces to
an ideal Maxwell-Boltzmann gas of nucleons.
In intermediate regions, the nucleus will
break up into clusters of various sizes and the system has some
features similar to that of a Bose gas.
More importantly, this model exhibits a phase transition
in the infinite $A$ but finite density limit with critical
parameters similar to other models of nuclear fragmentation.
As in percolation theory, the infinite cluster appears suddenly at the
critical point.  As in a Bose gas or Feynman's model of the $\lambda$
transition in liquid helium, the specific heat has a maximum at
the critical point with a discontinuous derivative.  The ``cusp''
in this model however is smoothed out by the finite size of the system.

This work supported in part by the National Science Foundation
Grant No. NSFPHY 92-12016.


\begin{figure}
\caption{The specific heat (a) and equation of state(b).  The specific
heat is Bose-like except at low temperatures, where it transitions to
Fermi gas like behavior.  In this region, the equation of state shows
the characteristic van de Waals behavior.}
\label{fig:thermodynamic-functions}
\end{figure}

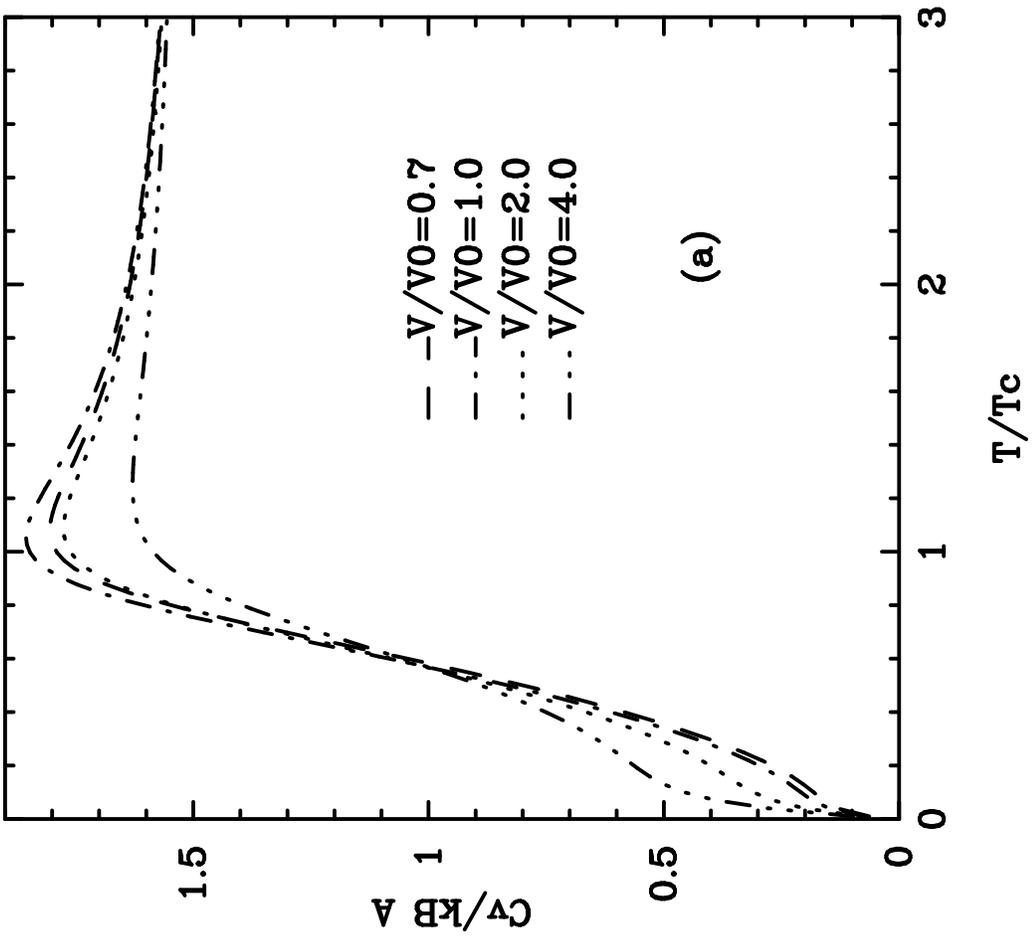
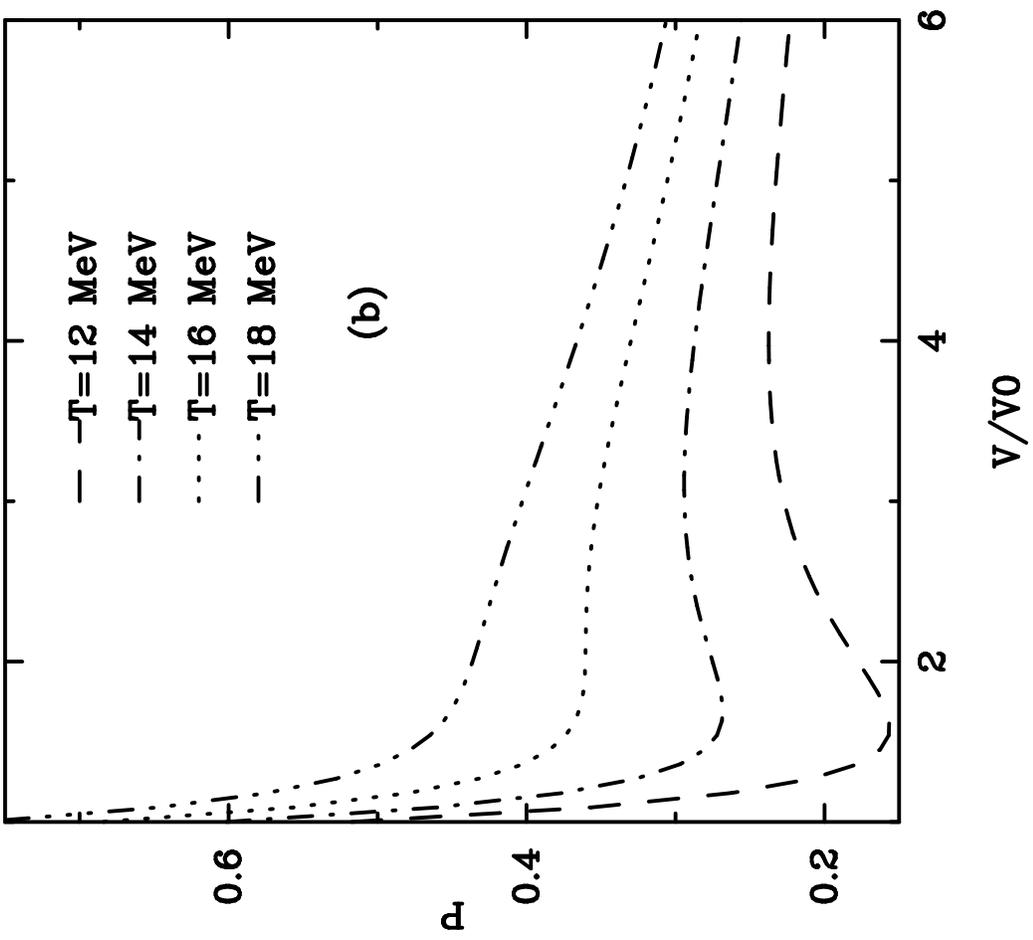
\end{document}